\documentclass{aa}  
\usepackage[varg]{txfonts}
\usepackage{graphics}
\usepackage{tikz}
\usepackage{epstopdf}
\begin{document}

   \title{Comparing focal plane wavefront control techniques:\\Numerical simulations and laboratory experiments}

   \author{A. Potier\inst{1}
          \and
          P. Baudoz\inst{1}
          \and
          R. Galicher\inst{1}
          \and
          G. Singh\inst{1}
          \and
          A. Boccaletti\inst{1}}

   \institute{\inst{1} LESIA, Observatoire de Paris, Université PSL, CNRS, Sorbonne Université, Université de Paris, 5 place Jules Janssen, 92195 Meudon, France\\
              \email{axel.potier@obspm.fr}
             }
   \date{Received 29 October 2019 / Accepted 4 March 2020}

 
  \abstract
   {Fewer than 1\% of all exoplanets detected to date have been characterized on the basis of spectroscopic observations of their atmosphere. Unlike indirect methods, high-contrast imaging offers access to atmospheric signatures by separating the light of a faint off-axis source from that of its parent star. Forthcoming space facilities, such as WFIRST/LUVOIR/HabEX, are expected to use coronagraphic instruments capable of imaging and spectroscopy in order to understand the physical properties of remote worlds. The primary technological challenge that drives the design of these instruments involves the precision control of wavefront phase and amplitude errors. To suppress the stellar intensity to acceptable levels, it is necessary to reduce phase aberrations to less than several picometers across the pupil of the telescope.}
   {Several focal plane wavefront sensing and control techniques have been proposed and demonstrated in laboratory to achieve the required accuracy. However, these techniques have never been tested and compared under the same laboratory conditions. This paper compares two of these techniques in a closed loop in visible light: the pair-wise (PW) associated with electric field conjugation (EFC) and self-coherent camera (SCC).}
   {We first ran numerical simulations to optimize PW wavefront sensing and to predict the performance of a coronagraphic instrument with PW associated to EFC wavefront control, assuming modeling errors for both PW and EFC. Then we implemented the techniques on a laboratory testbed. We introduced known aberrations into the system and compared the wavefront sensing using both PW and SCC. The speckle intensity in the coronagraphic image was then minimized using PW+EFC and SCC independently.}
   {We demonstrate that both techniques -- SCC, based on spatial modulation of the speckle intensity using an empirical model of the instrument, and PW, based on temporal modulation using a synthetic model -- can estimate the wavefront errors with the same precision. We also demonstrate that both SCC and PW+EFC can generate a dark hole in space-like conditions in a few iterations. Both techniques reach the current limitation of our laboratory bench and provide coronagraphic contrast levels of~$\sim 5.10^{-9}$ in a narrow spectral band (<0.25\% bandwidth).}
   {Our results indicate that both techniques are mature enough to be implemented in future space telescopes equipped with deformable mirrors for high-contrast imaging of exoplanets.}

   \keywords{Exoplanets --
                High-contrast imaging --
                Wavefront Sensor --
                Wavefront Control 
               }
   \titlerunning{Comparing focal plane wavefront control techniques}
   \maketitle
\section{Introduction}
By 2020, more than 4,000 exoplanets have already been discovered, mainly using   indirect detection techniques like transit or radial velocity. A few exoplanet atmospheres were probed using transit, high resolution spectroscopy, interferometry, and imaging. The transit method is used for planets that orbit at less than~$\sim1$ astronomical unit (AU) from their star \citep{essen19,espinoza19}. High-resolution spectroscopy of non-transiting planets \citep{snellen10,alonso19} and interferometry \citep{lacour19} currently focus on known exoplanets but without the ability to identify them. Imaging techniques to discover and spectrally characterize exoplanets in the outer part of the system have been devised \citep{Macintosh2015,Beuzit2019,konopacky13} and are planned for implementation in future space missions \citep{Debes19}. 

Imaging remains, nonetheless, challenging because exoplanets are 10$^4$ to 10$^{10}$ times fainter than their stars in visible and infrared light and they are separated from their star by a fraction of an arcsecond. The high-contrast imaging (HCI) community uses coronagraphs to attenuate the starlight and large telescopes equipped with adaptive optics (AO) systems to reach the required angular resolution. These technologies have been implemented on the current instruments such as the Spectro-Polarimetric High-contrast Exoplanet REsearch \citep[SPHERE,][]{Beuzit2019} at the very large telescope  and the Gemini Planet Imager \citep[GPI,][]{Macintosh2015} at the Gemini South observatory. These instruments are capable of discovering warm and young self-luminous exoplanets orbiting relatively far from their stars \citep[$\beta$ Pictoris b being one of the closest at 8 AU,][]{Lagrange2010} but they cannot detect fainter (mature or smaller) planets closer to their stars because of instrumental limitations. Indeed, current AO systems minimize the phase aberrations measured in the wavefront sensing channel but leave non-common path aberrations (NCPAs) in the science channel. Because of NCPAs, part of the stellar light goes through the coronagraphic imaging channel and induces stellar speckles on the science detector. In a similar manner, space-based telescopes are affected by slowly evolving aberrations which also create speckles in the science image \citep{Racine1999, Guyon2004, Martinez2012}. In order to detect fainter exoplanets, such aberrations must be minimized to a level of a few picometers rms over the pupil.

Thus, an active minimization of the stellar speckle intensity in the coronagraphic image is mandatory for the new generation of HCI instruments. The active control involves a focal plane wavefront sensor (FP~WFS) that measures the aberrations from the science image and a controller that drives deformable mirrors (DMs). Such a strategy will be used for the coronagraphic instrument on-board WFIRST \citep{mennesson18}.

The FP~WFS can use spatial modulations of the speckle intensity as performed by the self coherent camera \citep[SCC,][]{Baudoz2006,Mazoyer2013,Delorme2016Apr} or the asymmetric pupil Fourier wavefront sensor \citep{Pope2014}. Other techniques use temporal modulations of the speckle intensity, either in a small aberration regime like Pair-Wise probing \citep[PW,][]{Borde2006,GiveOn2007} or in a high aberration regime such as COFFEE \citep{Sauvage2012,Paul2013,Herscovici2018}.
Once the wavefront is measured, a wavefront controller (WFC) is needed to drive DMs. Several techniques have been proposed to find the optimal DM shape for a given estimated wavefront. The Energy Minimization algorithm minimizes the total energy of the speckle field in the region of interest called the Dark Hole (DH) where the exoplanets are searched \citep{Malbet1995,Borde2006}. The Electric Field Conjugation technique (EFC) derives a DM setting required to achieve a desired electric field in the focal plane \citep{GiveOn2007}. The performance of these techniques can be improved using regularization terms to account for example for the obstructed apertures or the use of two DMs \citep{Pueyo2009,Mazoyer2018} or in the case of large aberrations \mbox{\citep{Herscovici2018SPIE}}.
All these techniques (WFS and WFC) have been developed and tested independently in laboratories in different environmental conditions \citep{Mazoyer2019}. However, to our knowledge, none of them have been compared on the same testbed in a closed loop so far.\\

This paper compares the combination of PW and EFC with the SCC on the {\it tr\`es haute dynamique} (THD2) bench at the Paris Observatory. In Section~\ref{sec:WavefrontSensor}, we detail the theory behind PW and SCC WFS techniques and we also study the implementation and the robustness of~PW. In Section~\ref{sec:WavefrontControl}, two ways of controlling the wavefront aberrations are described: SCC and EFC. As both PW and EFC require an optical model of the instrument, a robustness study of the speckle minimization by PW+EFC is carried out. In Section~\ref{sec:THD2bench}, we present the THD2 bench, followed by the implementation of the combination PW+EFC, on one hand, and the SCC, on the other hand, in the laboratory. We measure and compare the wavefront aberrations and the contrast levels reached using each technique. We conclude the study in Section~\ref{sec:Discussion} with a discussion of the results obtained on the THD2 testbed, along with a listing of the pros and cons of the two techniques: SCC and PW+EFC.

\section{Wavefront sensors}
\label{sec:WavefrontSensor}
This section describes the principle behind the two FP~WFSs studied in this paper: the~SCC and the~PW. Both techniques measure the electric field in the science coronagraphic detector plane in a small aberration regime.

        \subsection{Model of light propagation}
        \label{subsec:ModelPropagation}
We model the light propagation inside a coronagraphic instrument. We call~$E_S$ the star electric field on the science detector. We express this field as a function of $\alpha$ and~$\beta$, the log-amplitude and phase aberrations in the pupil plane upstream of the coronagraphic mask:
\begin{equation}
\label{eq:totalelectricfield}
E_S=C[A\,e^{\alpha+i\beta}e^{i\phi}],
\end{equation}
where $A$ is the electric field in the pupil plane free from aberrations and, $\phi$ is the phase introduced by a DM settled in the pupil plane upstream of the coronagraphic mask. $C$ is the coronagraph linear operator that transforms the complex electric field from the pupil plane to the focal plane (science detector). Assuming a non-resolved star, the stellar light goes through the entrance pupil and is diffracted by a focal plane stellar coronagraph. The residual starlight is stopped by a Lyot-stop in the following conjugate pupil plane. Therefore, assuming Fourier optics, $C$ can be written as:
\begin{equation}
\label{eq:propmodel}
\begin{aligned}
C(E)= &\mathcal{F}\left[\mathcal{F}^{-1} \left[\text{M}\times \mathcal{F}(E)\right]\times \text{L}\right]\\
=& \left[\text{M}\times\mathcal{F}(E)\right]*\mathcal{F}(L),
\end{aligned}
\end{equation}
where $\mathcal{F}$ denotes the Fourier transform~(FT) operator, M represents the focal plane mask (FPM), and L is the classical binary Lyot stop. In the presence of aberrations, part of the stellar light goes through the system and reaches the science detector where stellar speckles are induced as a result. In case of small aberrations and small deformations of the DM, we can write the Taylor expansion of Eq.~\ref{eq:totalelectricfield} as:
\begin{equation}
\label{eq:totalelectricfield2}
E_S=C\left[Ae^{\alpha+i\beta}\right]+iC\left[A\phi\right]= E_{S_{0}}+E_{DM}.
\end{equation}
The field~$E_{S_{0}}$ is associated to the stellar speckles that are in the science image downstream the coronagraph because of the unknown upstream aberrations~$\alpha$ and~$\beta$. The field~$E_{DM}$ is associated to the star speckles that can be induced thanks to the~DM to compensate for~$E_{S_0}$ and therefore, to minimize~$E_S$ or its modulus. Before the minimization, one needs to measure the electric field~$E_{S_0}$. As the detector measures the intensity in visible and near-infrared light, we can only access the squared modulus of~$E_{S_0}$ in the science image. To retrieve the field from its modulus, FP~WFSs such as the SCC (Section~\ref{subsec:SCCWS}) or the PW (Section~\ref{subsec:PWWS}) modulate, respectively, the speckle intensity~$|E_{S_0}|^2$ either spatially or temporally.

        \subsection{The self-coherent camera}
        \label{subsec:SCCWS}
The SCC estimates the focal plane field from a spatial modulation of the speckle intensity. A small pinhole set next to the classical Lyot stop selects part of the starlight rejected by the FPM to create a reference channel \citep{Galicher2010,Mazoyer2013}. The residual starlight that propagates through this channel can interfere with the starlight that goes through the Lyot stop. The two fields recombine on the detector resulting in~$E_{SCC,}$
\begin{equation}
\label{eq:ElectricFieldSCC}
E_{SCC}(\overrightarrow{\eta})=E_{S_0}(\overrightarrow{\eta})+E_R(\overrightarrow{\eta})\text{exp}\left(\frac{-2i\pi\overrightarrow{\eta}\cdot\overrightarrow{\xi}}{\lambda}\right),
\end{equation}
where $E_R$ is the field induced by the light passing through the reference channel. $E_{S_0}$ is defined by~Eq.~\ref{eq:totalelectricfield}, considering~$\phi=0$ because no~DM phase is added in the beginning. The vectors~$\overrightarrow{\eta}$ and $\overrightarrow{\xi}$  describe the focal plane coordinates and the distance between the classical Lyot stop and the~SCC reference pinhole in the Lyot stop plane, respectively. In monochromatic light at wavelength~$\lambda$, the total intensity on the detector when using SCC can be written as:
\begin{equation}
\label{eq:i_scc}
I(\overrightarrow{\eta})=|E_{S_0}(\overrightarrow{\eta})|^2+|E_R(\overrightarrow{\eta})|^2+2\Re\left[E_{S_0}(\overrightarrow{\eta})E_R^*(\overrightarrow{\eta})\text{exp}\left(\frac{2i\pi\overrightarrow{\eta}\cdot\overrightarrow{\xi}}{\lambda}\right)\right].
\end{equation}
The first term is the speckle intensity that can be measured without~SCC. The second term is the~SCC reference channel intensity. The last term is the spatial modulation of~$E_{S_0}$ by the reference field~$E_R$. Once an intensity image~$I$ is recorded, its numerical inverse~FT can be calculated as:
\begin{equation}
\begin{aligned}
\mathcal{F}^{-1}[I](\overrightarrow{u}) & = \mathcal{F}^{-1}\left[|E_{S_0}|^2+|E_R|^2\right]*\delta\left(\overrightarrow{u}\right)\\
&+\mathcal{F}^{-1}\left[E_{S_0}\,E_R^*\right]*\delta\left(\overrightarrow{u}+\frac{\overrightarrow{\xi}}{\lambda}\right)\\
&+\mathcal{F}^{-1}\left[E_{S_0}^*\,E_R\right]*\delta\left(\overrightarrow{u}-\frac{\overrightarrow{\xi}}{\lambda}\right),
\end{aligned}
\end{equation}
where $\delta$ is the Dirac function and $\mathcal{F}^{-1}[I](\overrightarrow{u})$ is the inverse~FT of function~$I$ at the pupil plane position~$\overrightarrow{u}$. This~FT is composed of three peaks which do not overlap if the separation $\overrightarrow{\xi}$ between the classical Lyot stop and the~SCC reference pinhole is large enough. In such a case, one can isolate the lateral peak centered on~$\overrightarrow{u}=-\overrightarrow{\xi}/\lambda$ and call it $\mathcal{F}^{-1}[I_{\text{shifted}}]$\footnote{Any one of the lateral peaks can be selected because the two peaks are simply complex conjugates.} such that,
\begin{equation}
\mathcal{F}^{-1}[I_{\text{shifted}}]*\delta\left(\overrightarrow{u}+\frac{\overrightarrow{\xi}}{\lambda}\right)=\mathcal{F}^{-1}\left[E_SE_R^*\right]*\delta\left(\overrightarrow{u}+\frac{\overrightarrow{\xi}}{\lambda}\right).
\end{equation}
After centering the extracted peak, a second numerical FT results in
\begin{equation}
\label{eq:Imoins}
I_{\text{shifted}}=E_{S_0}E_R^* .
\end{equation}
Thus, by applying two numerical Fourier transforms on the recorded image~$I$, the electric field~$E_{S_0}$ of the stellar speckles present in the science image can be estimated \citep{Mazoyer2014}.

        \subsection{Pair-wise probing}
        \label{subsec:PWWS}
                \subsubsection{Theory}
PW probing uses temporal modulations of the speckle intensity to retrieve~$E_{S_0}$ \citep{GiveOn2007SPIE}. Similar to phase diversity \citep{gonsalves82}, several intensity images are recorded after introducing known aberrations called probes in the optical path. These probes can be created in the pupil plane by applying known shapes on the~DM. Assuming a small probe phase~$\phi_m$ in~Eq.~\ref{eq:totalelectricfield2}, the intensity recorded by the science detector can be written as:
\begin{equation}
I_m=|E_{S_{0}}+iC[A\phi_m]|^2.
\end{equation}
For each probe phase~$\phi_m$, a pair of images $I_m^+$ and $I_m^-$ are recorded corresponding to probes~$\pm\phi_m$. Then the difference between these images is calculated:
\begin{equation}
\label{eq:diff}
I_m^+-I_m^-=4(\Re(E_{S_{0}})\Re(iC[A\phi_m])+\Im(E_{S_{0}})\Im(iC[A\phi_m])),
\end{equation}
where $\Re(E_{S_{0}})$ and $\Im(E_{S_{0}}),$ respectively, represent the real and imaginary parts of the complex electric field $E_{S_{0}}$.
Considering~$k$ probes,~Eq.~\ref{eq:diff} can also be written for each pixel of the science image with coordinates~$(i,j)$ as:
\begin{eqnarray}
\begin{bmatrix}
\label{eq:estimation} 
 I_1^+-I_1^- \\
. \\
. \\
. \\
I_k^+-I_k^- 
\end{bmatrix}_{(i,j)}=4
\begin{bmatrix}
\Re (iC[A\phi_1]) & \Im (iC[A\phi_1]) \\
. & . \\
. & . \\
. & . \\
\Re (iC[A\phi_k]) & \Im (iC[A\phi_k]) 
\end{bmatrix}_{(i,j)}
\begin{bmatrix}
 \Re (E_{S_{0}}) \\
 \Im (E_{S_{0}})
\end{bmatrix}_{(i,j)}.
\end{eqnarray}
In order to fully retrieve $E_{S_{0}}$ at pixel (i,j), at least two of the~$k$ probes, called $\phi_m$ and $\phi_n$, must obey
\begin{eqnarray}
\label{eq:EnoughDiversity}
\Re(iC[A\phi_m])\Im(iC[A\phi_n])-\Re(iC[A\phi_n])\Im(iC[A\phi_m])\ne 0.
\end{eqnarray}
This condition imposes that at least two of the probes induce different electric fields~$E_{DM}$ at a particular location~$(i,j)$. The values of~$n$ and~$m$ can vary from one pixel to the other.

For all pixels for which~Eq.~\ref{eq:EnoughDiversity} is true, ~Eq.~\ref{eq:estimation} can be inverted to estimate the real and imaginary parts of the electric field~$E_{S_{0}}$:
\begin{eqnarray}
\label{eq:estimation2}
\begin{bmatrix}
 \Re (E_{S_{0}}) \\
 \Im (E_{S_{0}})
\end{bmatrix}_{(i,j)}=\frac{1}{4}
\begin{bmatrix}
\Re (iC[A\phi_1]) & \Im (iC[A\phi_1]) \\
. & . \\
. & . \\
. & . \\
\Re (iC[A\phi_k]) & \Im (iC[A\phi_k])
\end{bmatrix}^{\dagger}_{(i,j)}
\begin{bmatrix}
 I_1^+-I_1^- \\
. \\
. \\
. \\
I_k^+-I_k^-
\end{bmatrix}_{(i,j)},
\end{eqnarray}
where $X^\dagger$ is the pseudo inverse of matrix~$X$ calculated by the singular value decomposition (SVD) method. To conclude, PW can be implemented as follows: 1) We choose the~$k$ probes to be applied on the DM ; 2) We record the images $I_m^+$ and $I_m^-$ on the science detector adding the probes~$\pm\phi_m$ on the~$DM$ ; 3) We use a numerical model of the instrument to estimate the electric field $E_{DM}=iC[A\phi_m]$ added on each pixel of the science image for each probe~$\phi_m$ ; 4) We apply Eq.~\ref{eq:estimation2} to estimate~$E_{S_{0}}$ at the desired pixels using the recorded images~$I_m^+$ and~$I_m^-$.

A trade-off is required while choosing the number~$k$. On the one hand, a large number~$k$ of probes will ensure that~Eq.~\ref{eq:EnoughDiversity} is true for all pixels of interest. On the other hand, this number should be minimized to prevent the astrophysical data being contaminated by the probes during the science acquisition.
For estimating the speckle field~$E_{S_0}$ from~Eq.~\ref{eq:EnoughDiversity}, it is clear that at least two probes corresponding to~$4$ images are needed. The choice of the probes,~$\phi_m$ , is therefore a key element for~PW. In Section~\ref{subsubsec:ProbeChoice}, we consider the case where two and three actuators are used as probes. We then study the robustness of~PW versus a model error in Section~\ref{subsubsec:Robustestimate}.

\section{Numerical simulation of pair-wise probing}
\subsection{Assumptions of numerical simulations}
\label{subsubsec:assumptions}
\begin{figure}[!ht]
  \centering
  \includegraphics[height=8cm]{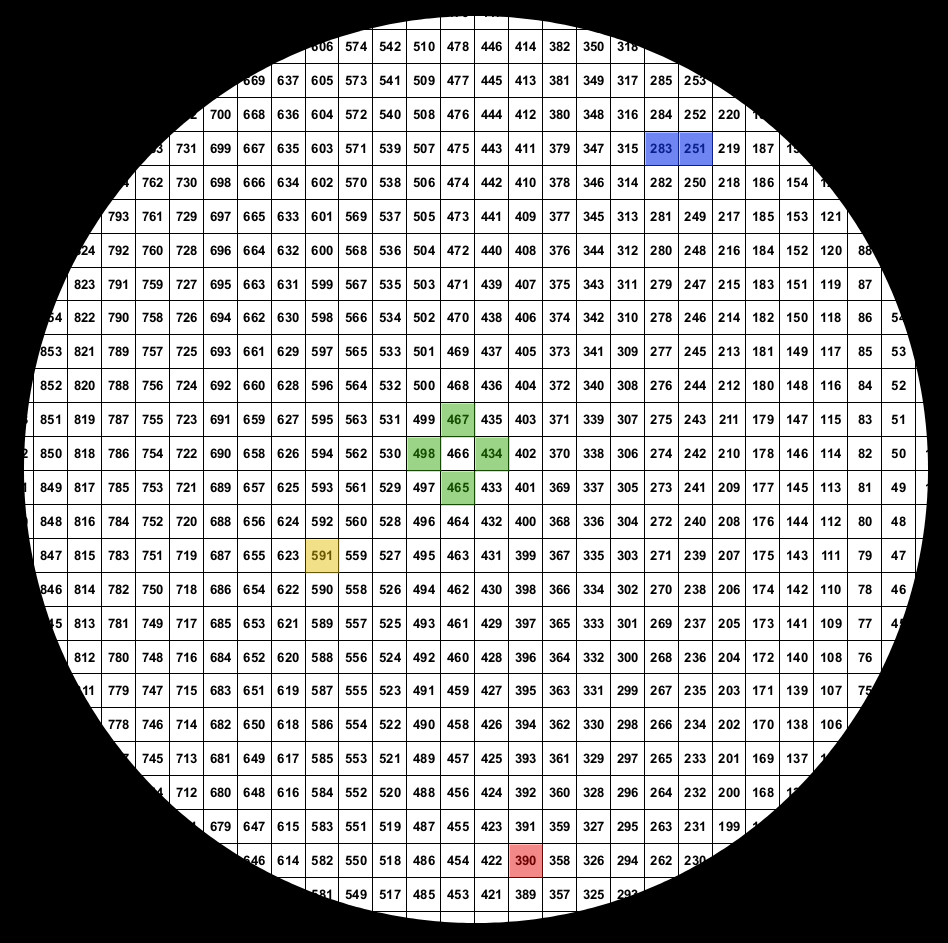}
   \caption{Position of DM actuators with respect to the pupil on the THD2 bench. The colored actuators correspond to different~PW probes tested in Sections \ref{subsubsec:ProbeChoice} and \ref{subsubsec:Robustestimate}. The association of actuator~$466$ (at DM center) with one of the actuators in green brings on a small error on the estimation of~$E_{S_0}$. The association of actuator~$466$ with the yellow actuator brings an average error. On the contrary, associating actuator~$466$ with the red actuator provides a bad estimation of~$E_{S_0}$ (see Section~\ref{subsubsec:ProbeChoice}). The blue actuators are used in Section~\ref{subsubsec:Robustestimate} to study the robustness of~PW in case of errors on the numerical model. }
   \label{fig:IndexDM}
   \end{figure}

The study in the following sections is based on the numerical simulations of the light propagation on the THD2 bench. Here, we briefly define a few simulation parameters (more detail in Section~\ref{subsec:THD2description}): a four-quadrant phase mask coronagraph \citep[FQPM,][]{Rouan2000} as a FPM, a science detector of 400x400 pixels with $7.55\,$pixels per resolution element, the position of the $28\times28$ actuators with respect to the pupil (see Fig.~\ref{fig:IndexDM}), and the influence function associated with each of the actuators \citep{Mazoyer2014SPIE}. We introduce a random phase aberration~$\beta$ with a power spectral density (PSD) proportional to the inverse of the spatial frequency to the power~$3$. Its standard deviation inside the pupil is~$20\,$nm. We also consider~$8\,\%~$rms error for the amplitude aberrations~$\alpha$ with a PSD proportional to the inverse of the square of the spatial frequency.

\citet{Matthews2017} and \citet{GiveOn2011} proposed to use $sinc$ functions as probes in the pupil plane to modulate the speckle intensity with a spatially uniform electric field in rectangular regions of the science image. We choose to use single actuator bumps because the phase induced when moving several actuators close to each other with a Boston Micromachine~DM can be non-linear with respect to the voltages because of the mechanical constrains. The influence function of each actuator is well constrained for the DM on THD2 \citep{Mazoyer2014SPIE}.

The choice of the bump amplitude is a trade-off. If it is too low, the signal from the difference $I_m^+-I_m^-$ stays below the noise level. If it is too high, the Taylor expansion of~Eq.~\ref{eq:totalelectricfield2} is no longer valid. We choose a peak-to-valley amplitude of $40\,$nm in numerical simulation.  We do not account for photon or detector noise.

\subsection{Probe choice: actuator bumps}
\label{subsubsec:ProbeChoice}

\subsubsection{Two pairs of probes}
We set actuator~$466$ as the first probe. This actuator is at the center of the~DM and also close to the center of the pupil. We then search for the second actuator that optimizes the~PW estimation in the case of two probes ($k=2$). We independently use each actuator located in the pupil as a second probe to estimate the electric field~$E_{S_0}$ defined in~Eq.~\ref{eq:estimation2}.

\begin{figure*}[!ht]
    \centering
   \includegraphics[width=18cm]{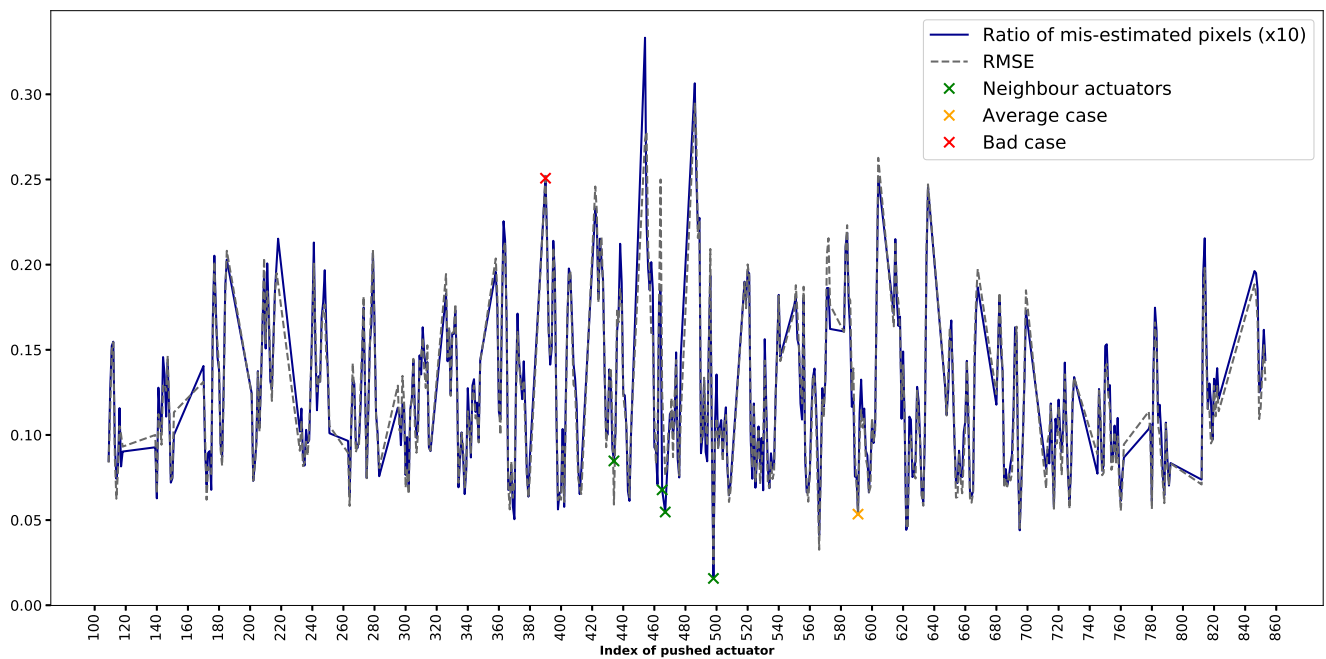}
   \caption[example] 
   { \label{fig:SSIM2act466} 
   Comparing the focal plane electric field with its PW estimate in the root mean square error metric (in dashed line in grey). The RMSE was calculated on the pixels where the difference between the true electric field and its focal plane estimate does not exceed three times the standard deviation of the true electric field. The pixels above this value are called "mis-estimated pixels". The ratio between the mis-estimated pixels and the total number of pixels in the DH area is multiplied by 10 and then plotted in blue. For the PW process, the first probe used is the actuator 466. The second probe is the bump of the actuator whose index is indicated in the x axis. The position of all these actuators are presented in Fig. \ref{fig:IndexDM}. The green crosses represent the ratio of mis-estimated pixels when the actuator 466 is associated with its four closest neighbors. The red cross represents a poor case when the actuator 466 is combined with the actuator 390. The actuator 591 is randomly chosen to illustrate an average result (yellow cross).}
\end{figure*}

To evaluate the quality of each estimation~$\hat{E}_{S_0}$, we first determine the true field,~$E_{S_0}$ , that is known in the numerical simulations and computed from~Eq.~\ref{eq:totalelectricfield} by equating~$\phi=0$. We calculate the standard deviation~$\sigma_0$ of~$E_{S_0}$ inside the DH of size $28\,\lambda/D\times28\,\lambda/D$ centered on the optical axis. For each estimation~$\hat{E}_{S_0}$, we calculate the root mean square error (RMSE) which is the average of $\sqrt{\left(\hat{E}_{S_0}-E_{S_0}\right)^2}$ over the~DH accounting only for pixels for which the difference~$|\hat{E}_{S_0}-E_{S_0}|$ is smaller than three times~$\sigma_0$.
This metric measures the accuracy of the estimation and checks if the estimation makes sense (smaller than~$3\,\sigma_0$). It is plotted in Fig.~\ref{fig:SSIM2act466} as a grey dashed line.
  
    We use a second metric that is the ratio of the number of mis-estimated pixels, that is,~for which the estimation error is larger than,~$3\,\sigma_0$, to the number of pixels inside the~DH. It is plotted in blue line in~Fig.~\ref{fig:SSIM2act466}). This metric measures the detector surface where the electric field is not adequately estimated.
    
These two metrics provide very similar results. As expected, the number of pixels where Eq.~\ref{eq:EnoughDiversity} is valid and the accuracy of the estimation of~$E_S$ are strongly correlated. We notice that the accuracy of the estimation is better when the second actuator comes closer to the first one (index~$466$) and is worse when it rolls away. We find that the best estimation according to both metrics is obtained for the actuator~$498$ that is one of the four closest neighbor of actuator~$466$ (see in Fig. \ref{fig:IndexDM}). In this case, the number of mis-estimated pixels is 0.16$\%$.

\begin{figure}[!ht]
   \centering
   \includegraphics[height=2.8cm]{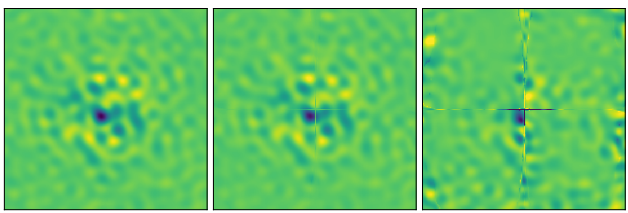}
   \caption[example] 
   { \label{fig:Images2act466_498} 
   Imaginary part of the true electric field (Left). Imaginary part of the estimated electric field with 466 and 498 actuators bumps as probes (Center). Ten times the difference between the two images (Right). The intensity scale is the same for all the images.}
   \end{figure}

Figure~\ref{fig:Images2act466_498} shows the imaginary part of the electric field~$E_S$ on the left, and its PW~estimation using a pair of actuators (466 and 498) in the center. The images are of size $28\,\lambda/D\times28\,\lambda/D$.
The difference between the two images multiplied by~$10$ is shown on the right of Fig.~\ref{fig:Images2act466_498}. The electric field is well estimated everywhere in the field of view except on the edge of the~DH and close to the FQPM transitions. For the latter, Eq.~\ref{eq:EnoughDiversity} is not valid because the light propagation model foresees a good extinction for the pixels along the FQPM transition whatever the pupil plane electric field is.

\begin{figure}[ht]
   \centering
   \includegraphics[height=8.3cm]{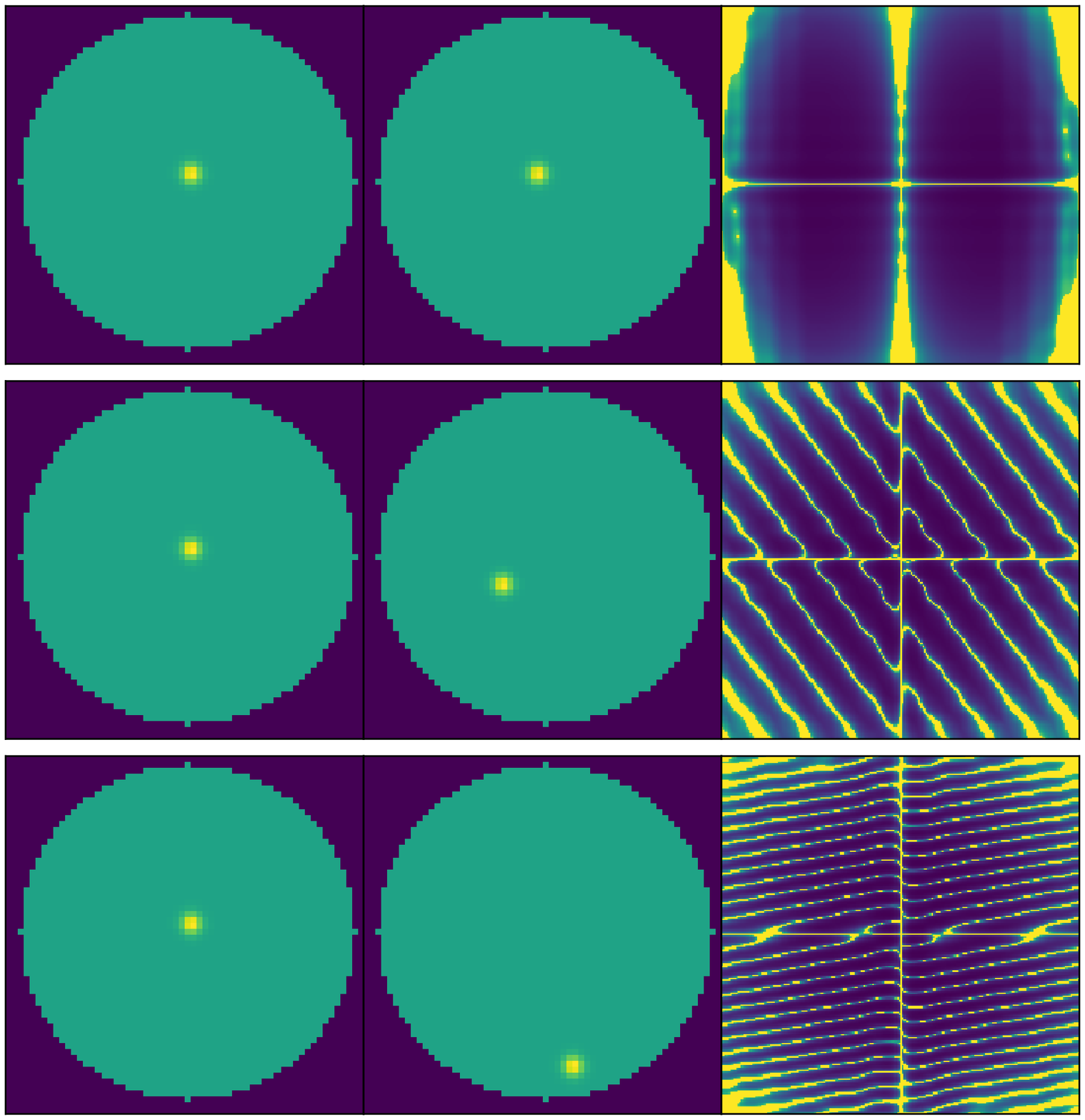}
   \caption[example] 
   { \label{fig:act466_498SVD15} 
   Top: positions of the 466 and 498 index actuators and their associated inverse eigenvalues (best case). Center: position of the 466 and 591 index actuators and their associated inverse eigenvalues (average case). Bottom: position of the 466 and 390 index actuators and their associated inverse eigenvalues (bad case). In the third column, 
   the highest of the inverse eigenvalues do not exceed the same threshold and appear bright.}
   \end{figure}

We used a third metric to verify the results obtained with the first two metrics. For a given pair of actuator-probes, we study the inverse of the singular values of the pseudo inverse matrix in~Eq.~\ref{eq:estimation2} at each pixel of the science detector. A high value indicates that the noise is enhanced and the estimation is not accurate. For a given DH, creating maps of these values is a practical tool to choose a pair of actuators. As an example, we show on the right of~Fig. \ref{fig:act466_498SVD15}, the maps for three different pairs of actuators whose positions are shown in the first two columns.

In these maps, the brighter are the pixels, the higher are the values and the poorer is the estimation of~$E_{S_0}$. If two actuators are close to each other, the inverse problem is well-posed in all the field of view except near the FQPM transitions and close to the edge of the DH. When the distance between actuators increases, the problem becomes ill-posed and periodic patterns of pixels where the estimation is inaccurate appear. The distribution of these pixels is important for a good estimation. For example, in an average case (yellow cross in Fig. \ref{fig:SSIM2act466} and middle row in~Fig.~\ref{fig:act466_498SVD15}) for which the RMSE and the ratio of misestimated pixels are low, the periodicity of misestimated pixels prevents the generation of a DH with a strong attenuation of the stellar speckles.
   
Therefore, the RMSE and the misestimated pixels ratio metrics alone are not accurate enough to determine a good probe combination. Using the map of the maximum of the inverse of the singular values appears to be an efficient complementary tool. In a future work, we will optimize this map to account for the distribution of the detector and photon noise that are not simulated here.

\subsubsection{Three probes}
We performed the same study using three probes instead of two. We tested all the triplets that included actuator 466. The best triplet slightly improves the estimation with respect to the case with two probes mainly at large angular separations in the final image. As already mentioned, one can increase the number of probes to obtain a better estimation of~$E_{S_0}$. However, during an actual observation, the choice between two or more probes will be driven by the time allocated for the speckle minimization versus the astrophysical observation.

\subsection{Robustness study}
\label{subsubsec:Robustestimate}
In Section \ref{subsubsec:ProbeChoice}, we assumed no error on the model of the instrument. This is not realistic because the thermal fluctuations and the changing mechanical flexures will always limit the precision of our knowledge on the state of the instrument. We note that the impact of model errors has been studied by \citet{Matthews2017} in the context of ground-based telescopes for~$sinc$ probes and for an apodized Lyot coronagraph. Here, we determine the impact of model errors assuming actuator probes and a perfect FQPM coronagraph in the space-like conditions (no atmospheric turbulence). We study three different cases: error on the influence function of the~DM actuator, translation, and rotation of the~DM relative to the pupil. For each case, we measure the RMSE for three~DH sizes discarding all pixels of the DH above~$3\,\sigma_0$ as explained in the previous section. The sizes of these DHs are $28\,\lambda/D \times 28\,\lambda/D$, $14\,\lambda/D \times 14\,\lambda/D$ and $7\,\lambda/D \times 7\,\lambda/D$. Under the assumptions described in Section~\ref{subsubsec:assumptions}, we first calculate the true electric field~$E_{S_0}$ followed by simulating the~PW technique using actuators~$466$ and~$498$ as probes.

\subsubsection{Influence function}
\begin{figure*}
   \centering
   \includegraphics[width=18cm]{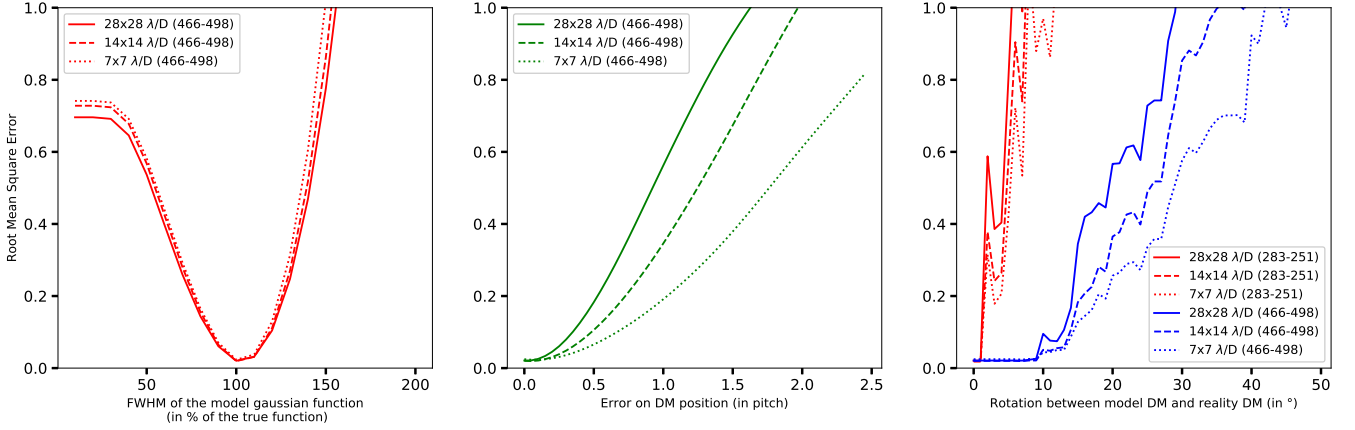}
   \caption{RMS error for three sizes of DH as a function of model errors. Left: influence function size error. Center: DM translation error. Right: DM rotation error for two different combinations of probes.}
   \label{fig:RobustnessEstimate}%
\end{figure*}

In this section, we study the impact of an error on the model of the influence function. First we simulate images~$I_m^\pm$ by considering the influence function of the actuators to be a Gaussian function with {\sc fwhm} equals to 1.2 times the pitch (the distance between two sequential actuators). When this {\sc fwhm} is used in the model, Eq.~\ref{eq:estimation2} provides the best estimation of ~$\hat{E}_{S_0}$. When we use a Gaussian function with a {\sc fwhm} in the model of the instrument that differs from the one used to simulate the images~$I_m^\pm$, the pseudo inverse matrix of~Eq.~\ref{eq:estimation2} deviates from the best solution. We test several {\sc fwhm} and, for each of them, we plot the RMSE metric on the left of~Fig.~\ref{fig:RobustnessEstimate}.
For the three~DH sizes, the~RMSE (error on the estimation) remains below $20\,\%$ as long as the error on the~{\sc fwhm} of the influence function stays below $25\%$. The knowledge of the influence function is therefore important for an accurate estimation of the electric field. To model the~DM on the~THD2 bench in the rest of the paper except in Section~\ref{subsec:RobustnessControl}, we use the non-Gaussian function measured by \cite{Mazoyer2014SPIE}.

\subsubsection{Actuator positions}
In this section, we study the impact of a model error on the~DM position. As previously, we acquire images,~$I_m^\pm$ , by fixing a certain position of the DM. This is referred to as the "true DM." 
We then consider that the modeled~DM array is translated relative to the true~DM in the horizontal direction in~Fig.~\ref{fig:IndexDM}. We calculate the estimated field,~$\hat{E}_{S_0}$ , and the corresponding~RMSE for each simulated error and for the three considered DHs. The central plot in~Fig.~\ref{fig:RobustnessEstimate} shows the RMSE results. A translation of the modeled~DM relative to the true one has more impact on the~PW estimation when the DH is larger. It means that the estimation of~$E_{S_0}$ is worst in the regions far from the optical axis. This is logical since the errors in the pupil plane are larger for higher spatial frequencies than for the lower spatial frequencies when the estimated aberrations in the pupil plane are translated relative to the true ones ($\alpha$ and~$\beta$). For instance, a translation error of one pitch implies a 20\% error in the 7$\lambda/D\times7\lambda/D$ region around the center, whereas it reaches 60\% in the largest DH (28$\lambda/D\times28\lambda/D$). The more uncertainties there are on the positions of the actuators, the narrower the region of correction.

We now consider that the modeled~DM is not translated relative to the true one but is rotated around the pupil center. We calculate the estimated field and the~RMSE for each rotation error. Results are plotted in blue on the right of~Fig.~\ref{fig:RobustnessEstimate}. As for the translation error, the larger the DH the more sensitive the~PW is to the rotation error. For example, the~RMSE is~10\% for the largest DH ($28\,\lambda/D \times 28\,\lambda/D$) and~5\% for the smallest one. Actuator~$498$ is at about two actuators from the pupil center such that a rotation of~$10^\circ$ corresponds to a translation of~$0.35\,$pitch. From the translation error plot (center plot of~Fig.~\ref{fig:RobustnessEstimate}), a~$0.35\,$pitch translation error gives a~RMSE of~$\sim10\%$ for the largest DH and~$\sim5\%$ for the smallest one. We therefore expect the~PW estimation to be more sensitive to an error on the angular position of the~DM if the actuator-probes are further away from the center of the pupil. We confirm this statement by executing the same study for a pair of actuator-probes closer to the edge of the pupil: actuators~$283$ and~$251$. The results plotted in red on the right of~Fig.~\ref{fig:RobustnessEstimate} confirm the following statement: the closer the pair of actuator-probes is to the center of the pupil, the more robust is the~PW with respect to a rotational error.

\section{Wavefront control}
\label{sec:WavefrontControl}
Once the electric field~$E_{S_0}$ is estimated, DMs are controlled to minimize the stellar speckle intensity inside a DH. In this section, we present two wavefront control techniques (SCC and EFC) assuming small aberrations ($\alpha$ and~$\beta$) and a single DM placed in the pupil plane. Both techniques use an iterative process and a control matrix.

    \subsection{SCC and EFC common strategy}
    \label{subsec:controlstrategy}
Classical AO systems measure and minimize the phase aberrations~$\beta$ in the pupil plane. In the context of HCI, this strategy is not optimal because the amplitude aberrations~$\alpha$ also induce stellar speckles in the science image. Moreover, DMs cannot control all the high spatial frequencies because of the limited number of actuators. Therefore, even if there are no amplitude aberrations, one~DM cannot completely null the phase~$\beta$. That is why \citet{Malbet1995} proposed to minimize the stellar speckle intensity inside a DH in the science image instead of the phase in a pupil plane. This has two main advantages. The field induced by both amplitude and phase aberrations can be minimized. And a stronger attenuation can be reached using the frequency-folding phenomenon and by decreasing the size of the DH \citep[DH,][]{Borde2006,GiveOn2006}.

We assume a single~DM placed in the pupil plane and we consider that the focal plane field $E_{DM}=iC[A\phi]$ induced on the science detector is a linear combination of the~DM actuator voltages~$\bar{a}$:
\begin{equation}
\label{eq:LinEFC}
E_{DM}=G\,\bar{a},
\end{equation}
where~$G$ is the linear transformation matrix between~$\bar{a}$ and~$E_{DM}$.

For the purposes of minimizing the speckle intensity, we search for the~DM voltages that minimize the electric field~$E_S=E_{S_0}+E_{DM}$ of~Eq.~$\ref{eq:totalelectricfield2}$ inside the~DH. In other terms, we minimize the following least mean squared criteria inside the~DH:
\begin{equation}
\label{eq:lms}
d^2=||E_{S_0}+G\,\bar{a}||^2.
\end{equation}
Several methods exist to solve this equation. We use a truncated~SVD to invert the matrix~$G$ and obtain the control matrix~$G^\dagger$. Indeed, the SVD is an easy-to-compute method to invert matrices and to minimize least-mean squared criteria. However, the problem is always ill-conditioned, which leads the derived solution $\bar{a}$ to be highly sensitive to any error in the computation of $G$ and $E_{S_0}$. Hence, we chose to regularize the SVD of $G$ by truncating the lowest singular values to decrease the condition number and to ensure a more stable solution.
Therefore, if we separate and concatenate the real and imaginary parts, one solution of~Eq.~\ref{eq:lms} can be written as:
\begin{equation}
\label{eq:actvoltage}
\bar{a}=-g[\Re (G)^\frown\Im (G)]^{\dagger}[\Re (E_{S_0})^\frown\Im (E_{S_0})] ,
\end{equation}
where $^\frown$ represents the concatenation. The field~$E_{S_0}$ is the one estimated by the FPWFSs such as the~SCC or~PW. Because of the linearization of~Eq.~ \ref{eq:totalelectricfield2}, we work in closed loop minimizing~$d$ in several iterations. The gain~$g$ ensures the loop convergence.

    \subsection{Control matrices implementation}
        \label{subsec:controlmatrices}
The main difficulty of the WFC strategy is to determine the matrix~$G$. In the case of SCC (Section~\ref{subsubsec:SCCMatrix}), we use an empirical matrix recorded prior to closing the correction loop. For the EFC (Section~\ref{subsubsec:EFCMatrix}), we use an analytical model of the instrument to calculate a synthetic matrix.

            \subsubsection{Self-coherent camera}
            \label{subsubsec:SCCMatrix}
The SCC technique does both focal plane wavefront sensing (Section~\ref{subsec:SCCWS}) as well as~WFC in closed loop. \citet{Mazoyer2014} showed that minimizing $I_{\text{shifted}}$ of~Eq.~\ref{eq:Imoins} is the same as minimizing~$E_{S_0}$ inside the~DH when the reference field $E_R$ is nonzero over the DH. This is the case in the configuration we test in Section~\ref{sec:THD2bench}. Therefore, we can replace~$E_{S_0}$ by~$I_{\text{shifted}}$ in~Eq.~\ref{eq:actvoltage}.

In the literature, the SCC interaction matrix~$G$ is an empirical matrix measured before applying the correction by recording SCC images while known sine and cosine patterns are applied on the~DM \citep{Poyneer2005}. For the $p$th sine/cosine function, $I_{\text{shifted\;in\;DH},p}$ is estimated from~Eq.~\ref{eq:Imoins}. Calling~$N$ the number of sine/cosine functions, the interaction matrix~$D$ gathers all the measurements
\begin{equation}
D=\begin{bmatrix}
 I_{\text{shifted\;in\;DH},1}\\
. \\
. \\
. \\
I_{\text{shifted\;in\;DH},N},
\end{bmatrix}
.\end{equation}
The~$G$ matrix can then be obtained using
\begin{equation}
G=D\,S,
\end{equation}
where~$S$ is the linear map between the~DM actuator voltages~$\bar{a}$ and the sine/cosine basis. 

            \subsubsection{Electric field conjugation}
            \label{subsubsec:EFCMatrix}
The second WFC that we study is the~EFC described in \citet{GiveOn2007}, also called speckle field nulling in \citet{Borde2006}. Unlike SCC, EFC is based on the model of the instrument. We take into account the same model which was used for the~PW (Section~\ref{subsec:PWWS}) to calculate the electric field~$E_{DM}$ induced by each actuator of the~DM inside the~DH. We decided to use the actuator basis \citep{Boyer1990} and note that a sine/cosine basis can also be implemented. We then calculated all the simulated fields to build the synthetic matrix~$G$. We can eventually use the synthetic matrix and the electric field~$E_{S_0}$ measured by~PW (Eq.~\ref{eq:totalelectricfield2}) to derive the~DM voltages from~Eq.~\ref{eq:actvoltage} to minimize the stellar speckle intensity inside the DH region.

The efficiency of~EFC as well as~PW is strongly correlated to the level of inaccuracy within the model. One can mitigate the impact of the inaccuracies truncating the SVDs. The~PW SVD is needed in~Eq.~\ref{eq:estimation2} for the wavefront sensing. In case of no truncation, the field~$E_{S_0}$ is accurately estimated everywhere in the~DH except at certain specific positions (bright areas in~Fig.~\ref{fig:act466_498SVD15}) that can induce bright speckles and lead to instabilities of the correction loop. If too many values are eliminated, the estimation of~$E_{S_0}$ is biased and the minimization is not effective. The~EFC SVD is needed in~Eq.~\ref{eq:actvoltage} for the~WFC. If no truncation is used then the noise and the estimation errors induce inaccurate motion of the~DM actuators. This will also lead to the instabilities of the correction loop. If too many values are truncated then the loop becomes stable but almost no modes are compensated by the~DM leading to no improvement in the speckle intensity minimization. In this work, we empirically chose the values of both PW and EFC truncations to obtain the best performance without diverging in numerical simulations. 

        \subsection{Robustness study of the PW+EFC closed loop}
        \label{subsec:RobustnessControl}
\begin{figure*}[htp]
   \centering
   \includegraphics[width=18cm]{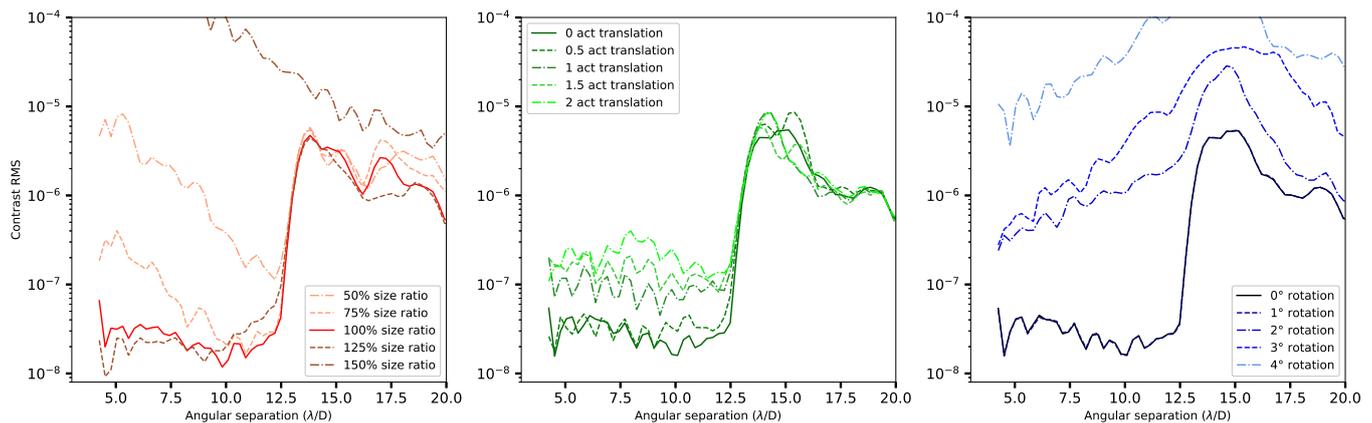}
   \caption[example] 
   { \label{fig:CorrectionRobustness} 
   RMS contrast as a function of the angular separation for different cases of model errors when implementing PW+EFC. Model errors simulated are the size of the influence function (left), a lateral translation of the DM (center), and a rotation of the DM (right).}
\end{figure*} 
This section presents the impact of model errors on the performance of the~PW+EFC correction loop. We consider the same errors as for~PW in Section~\ref{subsubsec:Robustestimate}: influence function size, translation and rotation of the~DM. We use the assumptions of Section~\ref{subsubsec:assumptions} except for the amplitude aberrations. Here they are assumed to be at 10$\%$ rms error and their PSD distribution is almost flat so that we may approach the testbed environment of the THD2. The two probes for the~PW technique are considered to be the bumps of the actuators~$466$ and~$498$. We fix the loop gain at~$g=0.5$.
We simulate~$578$ actuators in the pupil to calculate the matrix~$G$ and we select~$550\,$ modes after the~EFC SVD. As we study the correction of both amplitude and phase aberrations by a single~DM, the correction is done within a half DH spreading from~$2\,\lambda/D$ to $13\,\lambda/D$ on the horizontal axis and $-13\,\lambda/D$ to $13\,\lambda/D$ on the vertical axis. After the tenth iteration, the contrast level~$C$ is computed as the 1$\sigma$ azimutal standard deviation of the intensity in the coronagraphic science image divided by the maximum of the non-coronagraphic point spread function~(PSF). 

The results are shown in Fig \ref{fig:CorrectionRobustness}. In each plot, the full line is the performance with no model error. Model errors can strongly impact both the~WFS (Section~\ref{subsubsec:Robustestimate}) and the WFC.
In order to reach a~$10^{-7}$ contrast level, the size of the influence function has to be known with less than~$25\,\%$ error, the translation of the modeled~DM relative to the true one should be less than~$0.5\,$pitch and the orientation of the modeled~DM should be better than ~$1^\circ$.

\section{Lab performances: wavefront sensing and control on the THD2 bench}
\label{sec:THD2bench}
        \subsection{Bench description}
        \label{subsec:THD2description}
\begin{figure*}
   \centering
   \includegraphics[height=10cm]{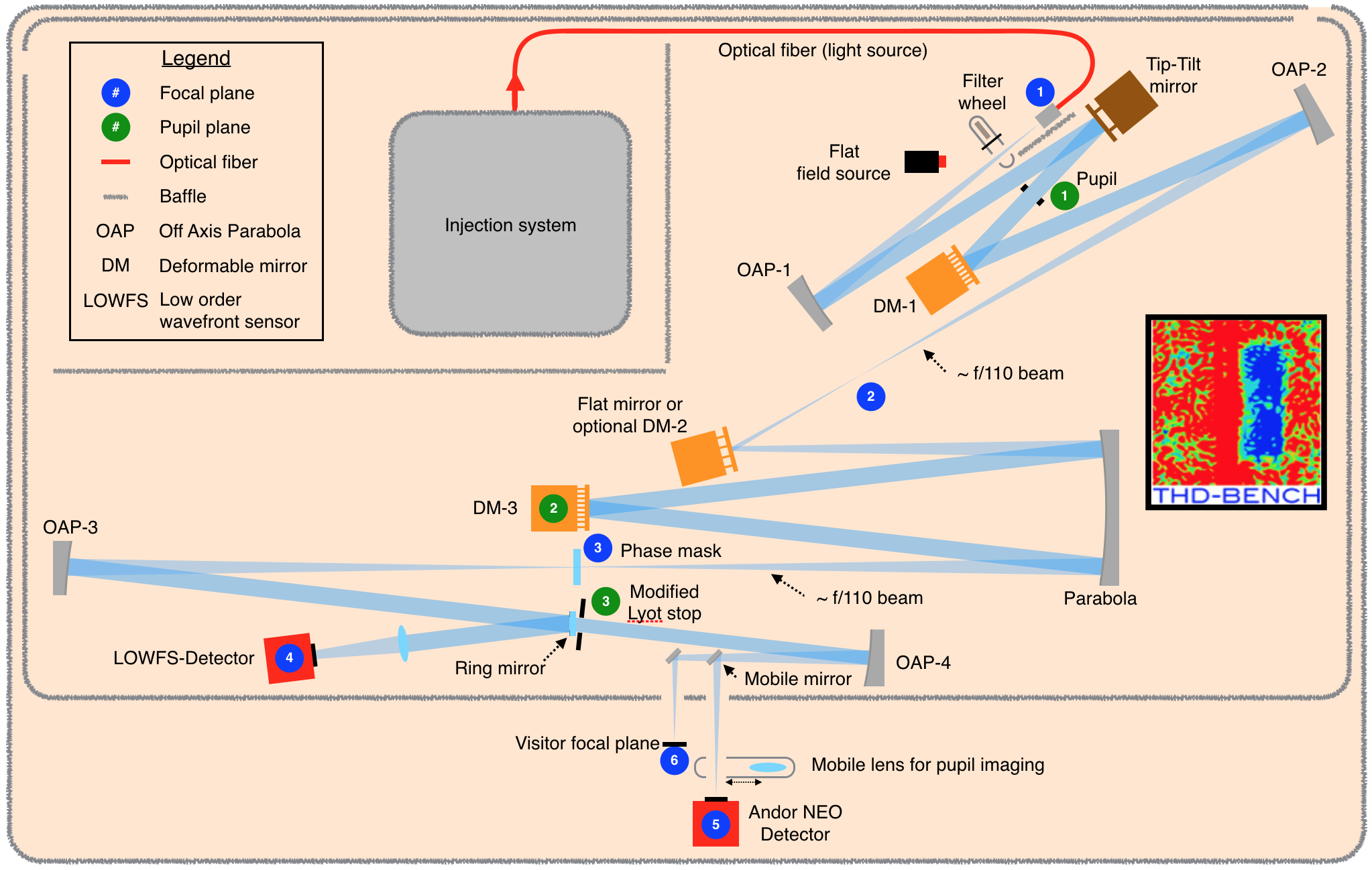}
   \caption{The layout of the THD2 bench presenting different optical components.}
              \label{fig:THD2}%
\end{figure*}

We compared the two wavefront sensing and control techniques described above on a HCI testbed developed at LESIA (Observatoire de Paris). The optical testbed, called THD2 for {\it tr\'es haute dynamique}, is located in an ISO7 pressurized clean room. It is described in detail in \cite{Baudoz2018} and its layout is shown in Fig.~\ref{fig:THD2}. In this paper, we used the following components:
\begin{itemize}
      \item An optical single mode fiber providing a monochromatic light source of wavelength $783.25\,$nm with a bandwidth less than $2\,$nm. The focal length of the first off-axis parabola (500 mm) flattens the Gaussian output of the fiber over the pupil diameter. The resulting amplitude aberration allows to reach a contrast level below $10^{-7}$ at $1\,\lambda/D$.
      \item An entrance pupil of $8.23\,$mm diameter.
      \item 32x32 Boston-Micromachine (DM3) settled in pupil plane~2.
      \item A FQPM located at the focal plane~3 in Fig.~\ref{fig:THD2}. 
      \item A Lyot-stop of $8$\,mm diameter in the pupil plane~3 (corresponding to a Lyot filtering of $97.2\,\%$). In this Lyot plane, a small pinhole with a diameter of~$0.4\,$mm is located at~$14.1\,$mm from the center of the Lyot stop and can be opened or closed to allow the use of the SCC. The ratio between the pinhole and the Lyot stop gives the first zero of the reference field at a radius of 24 $\,\lambda/D$. Thus, the SCC could theoretically correct a DH with a diameter up to 48 $\,\lambda/D$.
      \item Part of the light stopped by the Lyot-stop is reflected towards a Low Order Wavefront Sensor detector in the focal plane~4. This channel is used to stabilize the image of an on-axis star at the center of the~FQPM thanks to the tip-tilt mirror placed before the first pupil plane \citep{Singh2014}.
      \item A sCMOS camera recording images in the focal plane~5. 
\end{itemize}

The exact level of the phase induced by~DM3 was not well known because there is no absolute WFS on the THD2 bench. To calibrate~DM3, which is located in the pupil plane, we apply a cosine pattern with a small amplitude. This creates two copies of the PSF in the coronagraphic focal plane. By measuring the intensity of the copies with respect to the non coronagraphic PSF intensity, we infer the amplitude of the cosine optical path difference (OPD) that was introduced by the~DM. We then obtain the conversion factor from voltages to OPD.

The DM also has a non-linear response for each actuator that we numerically linearized using a quadratic function. Finally, neighbor actuators are coupled. However, for small displacements (less than~$100$\,nm), the relative accuracy on the actuator displacement is better than~$10\,\%$, which is not a limitation since we operate in closed loop. 

        \subsection{Wavefront sensors comparison}
        \label{subsec:WScomparison}
The PW technique is sensitive to the model errors, especially to the DM position with respect to the pupil (\ref{subsubsec:Robustestimate}).
 We took advantage of the previous implementation of the SCC on the THD2 bench to figure out the position of each actuators with respect to the pupil. We estimated that the actuators' positions with respect to the pupil are known with an accuracy of better than a~$0.2\,$pitch. The use of SCC for this measurement is not mandatory. It could be replaced by another WFS or pupil imaging. 

\begin{figure}[htp]
   \centering
   \begin{tikzpicture}
        \node (tiger) [anchor=south west, inner sep=0pt] {\includegraphics[width=8.3cm]{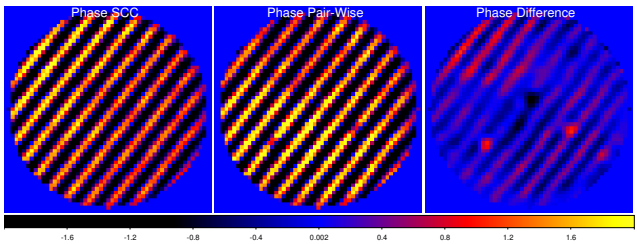}};
        \begin{scope}[x={(tiger.south east)},y={(tiger.north west)}]
        \fill[white] (7.8cm,0cm) rectangle (8.4cm,0.15cm);
    \end{scope}
    \begin{scope}[x={(tiger.south west)},y={(tiger.north east)}]
        \fill[white] (0cm,0cm) rectangle (0.5cm,0.15cm);
    \end{scope}
      \end{tikzpicture}
   \caption[example] 
   { \label{fig:PWCosshapeEstimation} 
   Left: SCC estimation (in nm) of an estimated 1.65 nm cosine. Center: PW estimation (in nm) of the same 1.65 nm cosine. Right: difference between the SCC estimation and 0.90 times the EFC estimation.}
   \end{figure}

We first used SCC and PW to retrieve a $1.65\pm0.05\,$nm cosine pattern that is applied to DM3. The SCC phase estimation is showed on the left of~Fig.~\ref{fig:PWCosshapeEstimation}.
   We then independently use PW using three probes: actuators numbered~$309$, $495,$ and~$659$ (Fig.~\ref{fig:IndexDM}) with an amplitude of~$33\pm3\,$nm. The electric field~$E_{S_0}$ is derived from~Eq.~\ref{eq:estimation2}. Finally, we use the inverse model of the instrument to get back to the pupil plane with a minor loss of information due to FQPM filtering \mbox{\citep{Mazoyer2013,Herscovici2018}}. The PW phase estimation is shown on the center of~Fig.~\ref{fig:PWCosshapeEstimation}. The right panel gives the difference between the SCC estimation and $0.90\,$times the PW estimation. The coefficient~$0.90$ was chosen to minimize the residuals. The location and orientation in the pupil plane of the cosine function are consistent for both methods. The~$10\,\%$ difference in phase amplitude might comes from the conversion from voltages to OPD that was calibrated with an accuracy of~$10\,\%$. This effect can easily be compensated during the correction by choosing a gain $g$ smaller than~$1$ in~Eq.~\ref{eq:actvoltage}.

\begin{figure}[htp]
    \centering
   \begin{tikzpicture}
        \node (tiger) [anchor=south west, inner sep=0pt] {\includegraphics[width=8.3cm]{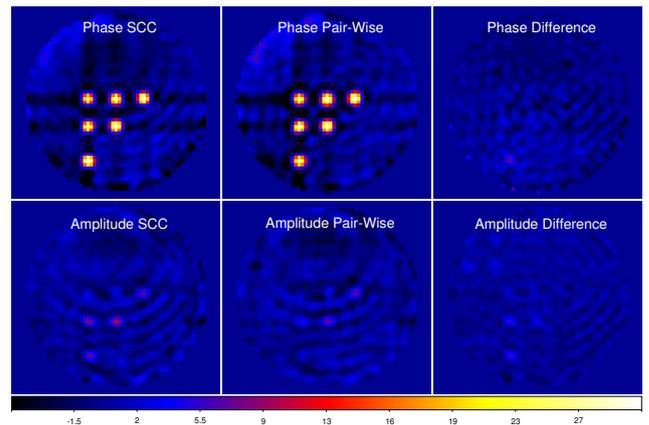}};
        \begin{scope}[x={(tiger.south east)},y={(tiger.north west)}]
        \fill[white] (7.8cm,0cm) rectangle (8.4cm,0.15cm);
    \end{scope}
    \begin{scope}[x={(tiger.south west)},y={(tiger.north east)}]
        \fill[white] (0cm,0cm) rectangle (0.5cm,0.15cm);
    \end{scope}
      \end{tikzpicture}
   \caption[example] 
   { \label{fig:PWFshapeEstimation} 
   Left: SCC estimation (in nm) of the 33 nm F-shape for both phase and amplitude aberrations. Center: PW estimation (in nm) of the 33 nm F-shape. Right: difference between the SCC estimation and 0.93 times the EFC estimation.}
   \end{figure}

We then used SCC and PW to retrieve a F-shape phase map induced by the poked actuators (six in total) with an amplitude of~$33\,$nm on~DM3. The estimated phase map (first row) and amplitude (second row) are shown in~Fig.~\ref{fig:PWFshapeEstimation} for SCC (left) and PW (center).
   The dark vertical and horizontal structures that are aligned with the poked actuators are the artifacts produced by the FQPM transitions which diffract light outside of the Lyot stop. We proved in a laboratory setting that this is not a limitation for efficient correction \citep{Mazoyer2013} inside a DH because we do not need to back propagate the phase and amplitude aberrations in the pupil plane. Here, we do the back propagation only for the purposes of presentation. We find that SCC and PW provide very similar phase estimation. The difference between the estimation of SCC and~$0.93$ times the estimation of PW is shown on the right of the figure. The coefficient~$0.93$ was again chosen to minimize the energy of the difference. As previously, the coefficient might come from the voltage-to-nanometer accuracy. The F-shape pattern is also detected in the amplitude images. This is because both SCC and PW measure the second order in the Taylor expansion of $e^{i\phi}$ with $\phi$ the phase induced by the DM. For a~$33\,$nm phase aberration at~$783.25\,$nm, we expect an amplitude aberration of $\phi^2/2\simeq3.5\,\%$. Converted to OPD unit at~$785.25\,$nm, we find an amplitude error of~$4.4\,$nm. In the SCC and PW amplitude estimation, we measure an amplitude error of~$4.2\,$nm rms for PW and $6.5\,$nm rms for SCC, which is consistent with what was expected. 

        \subsection{Wavefront control comparison}
        \label{subsec:WCcomparison}
\begin{figure*}[htp]
  \centering
   \begin{tikzpicture}
        \node (tiger) [anchor=south west, inner sep=0pt] {\includegraphics[width=18cm]{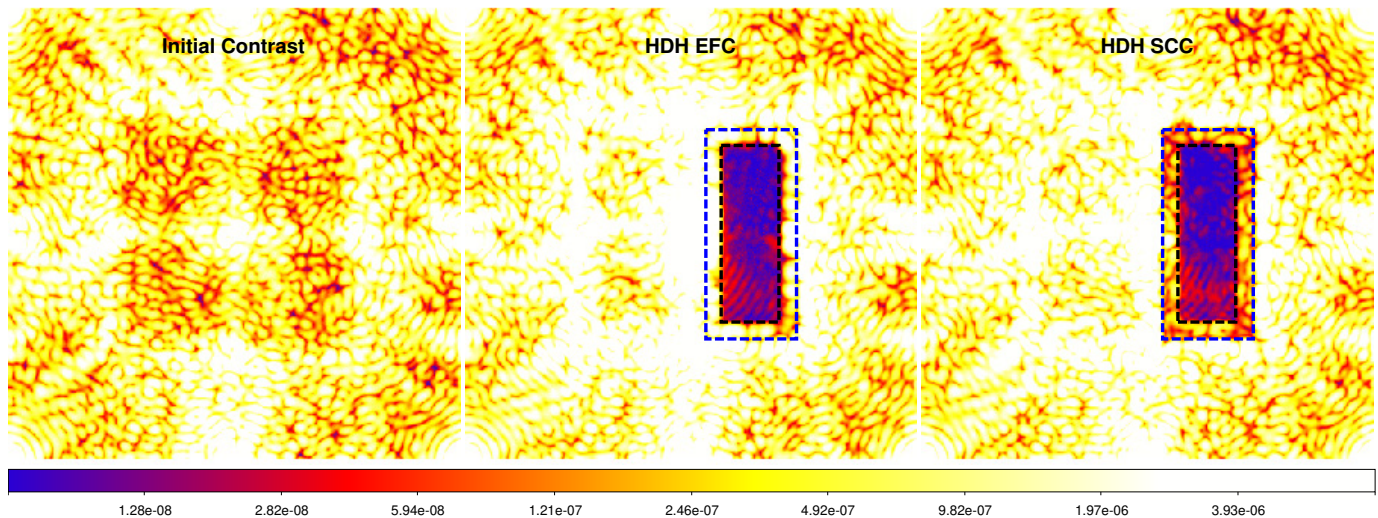}};
        \begin{scope}[x={(tiger.south east)},y={(tiger.north west)}]
        \fill[white] (17.5cm,0cm) rectangle (18cm,0.20cm);
    \end{scope}
    \begin{scope}[x={(tiger.south west)},y={(tiger.north east)}]
        \fill[white] (0cm,0cm) rectangle (0.5cm,0.20cm);
    \end{scope}
      \end{tikzpicture}
   \caption[example] 
   { \label{fig:Correction} 
   Left panel: initial raw contrast.  Center panel: Raw Half DH contrast correction done with the EFC.  Right panel: raw Half DH contrast correction done with the SCC. The blue rectangle corresponds to the region 2-13 $\lambda/D$ and -13-13$\lambda/D$ half DH. After ten iterations, the DH size is decreased to a size of $4\times11\lambda/D$ and $-11\times11\lambda/D$ shown here in the inner black rectangle.}
   \end{figure*}

\begin{figure}[htp]
   \centering
   \includegraphics[width=8.3cm]{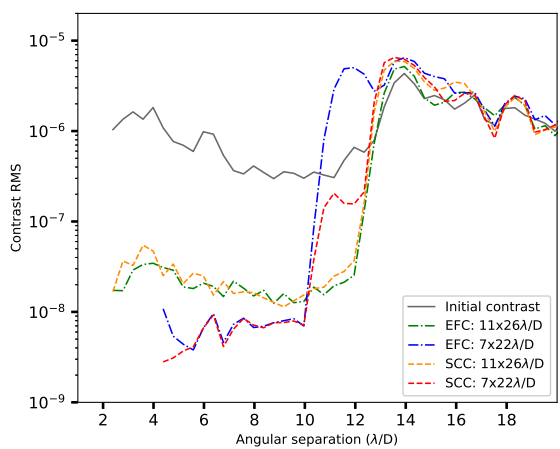}
   \caption[example] 
   { \label{fig:Contrast} Comparison of 1$\sigma$ contrast inside a half DH versus the angular separation obtained when implementing PW+EFC and the SCC on the THD2 bench.}
   \end{figure}

We finally tested the correction loop considering both SCC and PW+EFC independently. In both cases, we started from the same DM voltages implying that the same initial phase and amplitude aberrations were considered prior to testing. The initial image corresponds to a coronagraphic image where the phase was pre-corrected with SCC to reach a contrast level of $\approx10^{-6}$ (Fig. \ref{fig:Correction},~left).
As a first experiment, we used the PW technique to estimate the field as explained in Section~\ref{subsec:WScomparison} and EFC with~$550$ modes to create a half DH from $2\,\lambda/D$ to $13\,\lambda/D$ in one direction and from $-13\,\lambda/D$ to $13\,\lambda/D$ in the other direction. After ten iterations, we calculate the 1$\sigma$ contrast~$C$ inside the half DH as defined in Section~\ref{subsec:RobustnessControl}. We plot~$C$ against the function of the angular separation as a green dash-dot line in~Fig.~\ref{fig:Contrast}.
   The contrast remains between $10^{-8}$ and $3\times10^{-8}$ in the range~2-~$12\,\lambda/D$. This result is in very good agreement with the numerical simulation obtained with no model error (full lines in~Fig.\,\ref{fig:CorrectionRobustness}).
We conclude that our model of the THD2 bench is capable of providing a correction at a~$<10^{-8}$ contrast level accuracy. The same closed loop algorithm was also implemented with only two probes ($466$ and~$498$), which led to the same contrast level but it proved to be less stable with time. A complete study of EFC stability is in progress and out of the scope of the current paper.

Starting from the last iteration of the first experiment, the correction area was reduced by $2\,\lambda/D$ on each side to get a smaller half DH of size $7\,\lambda/D \times 22\,\lambda/D$. The correction reached a limit after ten iterations without diverging. The resulting image is presented in the center of~Fig.~\ref{fig:Correction} and the corresponding contrast is plotted in~Fig.~\ref{fig:Contrast} as a blue dash-dot line. The contrast reaches a level of $7.5\times10^{-9}$ at 1$1\sigma$ inside the DH. Consequently, we improved the contrast level by a factor of about~$2$ by decreasing the size of the DH. A pattern visible inside the DH (series of rings) seems to originate from a ghost reflection induced by the FQPM. These artifacts cannot be corrected because the light is not coherent with the central source and it sets a $8\times10^{-9}$ contrast limit in the lower part of the DH. In the top part, we measure an averaged 1$1\sigma$ contrast level of~$4\times10^{-9}$. 

On the same day, we used SCC starting from the same aberrations as assumed for the PW probing test (left image of~Fig.~\ref{fig:Correction}). The only difference in the settings is the presence of the SCC reference pinhole in the Lyot stop plane. An interaction matrix was recorded by applying sine/cosine functions on the DM3. We first minimize the speckle intensity inside the same region of the DH ($11\,\lambda/D\times 26\,\lambda/D$) as performed for the PW+EFC. After~$10$ iterations, we stop the loop. We close the reference channel and, we record the coronagraphic image. We then open the reference channel, change the DH size to $7\,\lambda/D\times 22\,\lambda/D$ and, apply the SCC correction for~$10$ iterations. The resulting coronographic image is presented on the right of~Fig.~\ref{fig:Correction}. The  contrast levels for the two DH sizes are plotted in~Fig.~\ref{fig:Contrast} in orange (larger DH) and red dashed lines. The detection of the ghost and the contrast level in all parts of the images are very similar to the ones obtained with the PW+EFC. Both the PW+EFC and the SCC techniques enable a similar minimization of the speckle intensity at a contrast level of~$\sim5\times10^{-9}$.

\section{Discussion}
\label{sec:Discussion}
This paper presents the first laboratory comparison of the two WFS/WFC algorithms: one which uses a spatial modulation of the speckle intensity (SCC) and the other based on the temporal modulation (PW+EFC). We list the pros and cons of both techniques in this section.

\subsection{Performance of WFS/WFC on THD2}

The SCC has already been demonstrated on the THD2 bench \citep{Baudoz2018SPIE}, where the contrast levels down to~$10^{-8}$ in the region $-12\,\lambda/D \times 12\,\lambda/D$ in a full DH are obtained by controlling two DMs (DM3 in the pupil plane and DM1 situated at~$26.9\,$cm from the pupil plane in a collimated beam). In this paper, we used only DM3 and reached similar performance (Section~\ref{subsec:WCcomparison}). However, we had to focus on half of the field of view because we used a single DM instead of two.

Apart from the optical ghost probably induced by the~FQPM, there are several other sources that prevent reaching the contrast level below~$5\times10^{-9}$ on THD2. First, the basic correction algorithm used for both EFC and SCC may drives the contrast level to a local minimum. Other minimization techniques based on the regularization terms \citep{Pueyo2009,Mazoyer2018,Herscovici2018SPIE} may be required to improve the minimization of the speckle intensity. This study is currently in progress in the laboratory. Moreover, the electronics of the DM3 that use a 14-bit ADC also limit the contrast above~$5\times10^{-9}$. It will be upgraded to 16-bit ADC in the coming months. In addition, the testbed is not under vacuum, so the internal turbulence may arise from thermal and mechanical variations. In that case, the SCC algorithm using a single image per loop would present an advantage over the PW+EFC, which is slower because it requires at least four images at each iteration.

        \subsection{Temporal versus spatial modulation}
In this section, we discuss the advantages and drawbacks of the two techniques implemented on the THD2 testbed, applied in space-like conditions and using only one DM for correction.

There is no significant difference in the contrast performance. Both techniques reach the current limits of the THD2 bench. Each iteration of the SCC technique requires a single image so the correction is faster than for the PW+EFC which need at least five images (four for the PW estimation and one for the astrophysical purpose). To sample the spatial fringes, the SCC however requires about three times more pixels on the science detector than the PW+EFC. When combined with the number of images per iteration, SCC spreads the light over less pixels though: three times more pixels per image but five times less images per iteration. However, \citet{Borde2006,GiveOn2007} proposed solutions to reduce the number of images for the PW+EFC.
The SCC reference beam adds a flat distribution of light that cannot mimic a planet signal. The downside is that it adds photon noise. It may thus be required to adjust the diameter of the reference channel during the observation so that the reference flux is always below the speckle level. The reference flux can also be used a posteriori for coherence differential imaging because it spatially modulates the stellar speckles that remains after WFC \citep{Baudoz2012}. We can imagine that the probe images of the PW technique can also be used for the coherence differential imaging.

Up until now, the SCC has used an empirical interaction matrix that has to be recorded before the correction loop is closed. The matrix can be impacted by the detection noise and, above all, it requires telescope time (at least the instrument time if the matrix is recorded using the internal source). This strategy is not optimal if the matrix needs to be updated regularly to account for the changes in the instrument configuration. On the contrary, PW+EFC use a numerical model of the instrument so that several synthetic interaction matrices can be calculated before the observations for numerous instrument configurations. These matrices are, however, very sensitive to model errors. In both cases (empirical and synthetic matrix), the matrices can be useless if one parameter of the instrument suddenly changes. That is why, a semi-empirical solution may be required: regular recording of a few data to modify the synthetic or empirical matrix. Our team is currently investigating such solutions.

The PW+EFC combination can easily be implemented in any coronagraphic instrument that includes a DM. On the contrary, the SCC requires optics large enough to allow the light of the reference pinhole to propagate from the Lyot stop plane to the final detector. This condition is not a strong drawback for future instruments but it prevents the implementation of the SCC on most of the current instruments which were not designed with such a flexibility.

The current versions of the SCC and the PW+EFC which are implemented on the THD2 bench use a basic truncated SVD to calculate the control matrix. More advanced solutions adding regulation terms for example may help to improve the stability and the performance of both techniques.

\section{Conclusion}
This paper described and compared two high-contrast imaging techniques. Both techniques retrieve the electric field associated to the stellar speckle in the science image and control DMs to minimize the speckle intensity.
One of the techniques, called the self-coherent camera (SCC), uses spatial modulations of the speckle intensity and an empirical model of the instrument. The other, pair-wise probing associated with electric field conjugation (PW+EFC), is based on temporal modulations and a synthetic model.

We first provided a mathematical description of these techniques. Then we used numerical simulations to demonstrate that PW is more efficient if the two actuators used as probes are close to each other. In simulations, we also studied the robustness of PW as well as PW+EFC when model errors, such as the knowledge on the DM position and influence function, are taken into account. We finally demonstrated and compared the two techniques in laboratory on the THD2 bench. We tested the SCC and PW under the same phase and amplitude aberrations to show that both techniques were capable of measuring the aberrations with a subnanometer accuracy. We compared PW+EFC and SCC abilities to generate a dark hole in space-like conditions in a few iterations. Both techniques converge to a contrast of $\sim5\times10^{-9}$ between~$2\,\lambda/D$ and~$12\,\lambda/D$ and are mainly limited by an optical ghost. In this paper, both techniques were studied and compared in monochromatic light. It can also be done in broadband using hardware or software upgrades for both SCC \citep{Delorme2016Apr} and PW+EFC \citep{Seo17}.

We discussed the advantages and drawbacks of each technique. In terms of wavefront sensing and control, both techniques provide similar performance down to~$5\times10^{-9}$ contrast levels. One of the advantage of the SCC is that it enables coherence differential imaging that can improve, a posteriori, the contrast achieved after the active minimization of the speckle field. It is, however, more complicated to implement it on the existing instruments than the PW+EFC. Our main conclusion is that the two techniques are mature enough to be implemented in future space telescopes equipped with DMs for high-contrast imaging.
Future studies are planned to include testing these techniques in more realistic configurations with obstructed pupils and broadband imaging using several DMs in cascade.

\begin{acknowledgements}
The authors thank the Centre National d'Études Spatiales (CNES) which co-funds Axel Potier PhD work. We particularly thank Jean-Michel Le Duigou for his support and Johan Mazoyer for his comments. AP also thank the Île-de-France region, and its programm DIM-ACAV+ for being the second co-funder of its PhD work. GS would like to acknowledge the funding received from the European Union's Horizon 2020 research and innovation programme under the Marie Sklodowska-Curie grant agreement No 798909. This work was carried out as part of IRIS Origines et Conditions d’Apparition de la Vie funded by Idex with reference ANR-10-IDEX-0001-02 PSL*.
\end{acknowledgements}

\bibliographystyle{aa} 
\bibliography{bib_GS}

\end{document}